\documentclass{article}
\usepackage{amssymb}
\usepackage{amsmath}
\usepackage{cite}

\begin{document}

\title{Noether's Theorem and Symmetry}
\author{
A.K. Halder \\
{\ \textit{Department of Mathematics, Pondicherry University, Kalapet 605014, India}}
\\
\and
Andronikos Paliathanasis\thanks{%
Email: anpaliat@phys.uoa.gr} \\
{\ \textit{Instituto de Ciencias F\'{\i}sicas y Matem\'{a}ticas, Universidad
Austral de Chile, Valdivia, Chile}}
\\
{\ \textit{Institute of Systems Science, Durban University of Technology
}}\\
{\ \textit{PO Box 1334, Durban 4000, Republic of South Africa}}
\\
\and 
P.G.L. Leach \\
{\ \textit{Institute of Systems Science, Durban University of Technology
}}\\
{\ \textit{PO Box 1334, Durban 4000, Republic of South Africa}}
\\
{\ \textit{School of Mathematical Sciences, University of KwaZulu-Natal
}}\\
{\ \textit{Durban,  Republic of South Africa}}}
\maketitle

\begin{abstract}
In Noether's original presentation of her celebrated theorm of 1918
allowance was made for the dependence of the coefficient functions of the
differential operator which generated the infinitesimal transformation of
the Action Integral upon the derivatives of the depenent variable(s), the
so-called generalised, or dynamical, symmetries. A similar allowance is to
be found in the variables of the boundary function, often termed a gauge
function by those who have not read the original paper. This generality was
lost after texts such as those of Courant and Hilbert or Lovelock and Rund
confined attention to point transformations only. In recent decades this
dimunition of the power of Noether's Theorem has been partly countered, in
particular in the review of Sarlet and Cantrijn. In this special issue we
emphasise the generality of Noether's Theorem in its original form and
explore the applicability of even more general coefficient functions by
alowing for nonlocal terms. We also look for the application of these more
general symmetries to problems in which parameters or parametric functions
have a more general dependence upon the independent variables.\newline
\end{abstract}

\section{Introduction}

Noether's Theorem \cite{Noether 18 a} treats the invariance of the
functional of the Calculus of Variations -- the Action Integral in Mechanics
-- under an infinitesimal transformation. This transformation can be
considered as being generated by a differential operator which in this case
is termed a Noether symmetry. The theorem was not developed \textit{ab initio%
} by Noether. Not only is it steeped in the philosophy of Lie's approach but
also it is based on earlier work of more immediate relevance by a number of
writers. Hamel \cite{Hamel 59, Hamel 50} and Herglotz \cite{Herglotz 36} had
already applied the ideas developed in her paper to some specific finite
groups. Fokker \cite{Fokker 17 a} did the same for specific infinite groups.
A then recently published paper by Kneser \cite{Knesner 18 a} discussed the
finding of invariants by a similar method. She also acknowledges the
contemporary work of Klein \cite{Klein 18 a}. Considering that the paper was
presented to the \textit{Festschrift} in honour of the fiftieth anniversary
of Klein's doctorate this final attribution must have been almost obligatory.

For reasons obscure Noether's Theorem has been subsequently subject to
downsizing by many authors of textbooks \cite{Rund 75 a, Hilbert 53 a, Logan
77 a} which has then given other writers (\textit{cf} \cite{Stan 94 a}) the
opportunity to `generalize' the theorem or to demonstrate the superiority of
some other method \cite{Ancos,Hojman 92 a} to obtain more general results
\cite{GonzalezGascon 94 a,cartan01,cartan02}. This is possibly due to the
simplified form presented in Courant and Hilbert \cite{Hilbert 53 a}. As
Hilbert was present at the presentation by Noether of her theorem to the
\textit{Festschrift} in honour of the fiftieth anniversary of Felix Klein's
doctorate, it could be assumed that his description would be accurate.
However, Hilbert's sole contribution to the text was his name.

This particularizing tendency has not been uniform, \textit{eg} the review
by Sarlet and Cantrijn \cite{Sarlet 81 a}. According to Noether \cite%
{Noether 18 a}[pp 236-237] `In den Transformationen k\"{o}nnen auch die
Ableitungen der $u$ nach den $x$, also $\partial u/\partial x$, $\partial
^{2}u/\partial x^{2}$, $\ldots $ auftreten' so that the introduction of
generalised transformations is made before the statement of the theorem \cite%
{Noether 18 a}[p 238]. On page 240, after the statement of the theorem,
Noether does mention particular results if one restricts the class of
transformations admitted and this may be the source of the usage of the
restricted treatments mentioned above.

We permit the coefficient functions of the generator of the infinitesimal
transformation to be of unspecified dependence subject to any requirement of
differentiability.

For the purposes of clarity of exposition we develop the theory of the
theorem in terms of a first-order Lagrangian in one dependent and one
independent variable. The expressions for more complicated situations are
given below in a convenient summary format.\hfill

\section{Noether Symmetries}

We consider the Action Integral
\begin{equation}
A=\int_{t_{0}}^{t_{1}}L\left( t,q,\dot{q}\right) \mbox{\rm d}t.  \label{4421}
\end{equation}%
Under the infinitesimal transformation
\begin{equation}
\bar{t}=t+\varepsilon \tau ,\qquad \bar{q}=q+\varepsilon \eta   \label{4422}
\end{equation}%
generated by the differential operator
\[
\Gamma =\tau \partial _{t}+\eta \partial _{q},
\]%
the Action Integral (\ref{4421}) becomes
\[
\bar{A}=\int_{\bar{t}_{0}}^{\bar{t}_{1}}L\left( \bar{t},\bar{q},\dot{\bar{q}}%
\right) \mbox{\rm d}\bar{t}
\]%
($\dot{\bar{q}}$ is $\mbox{\rm d}\bar{q}/\mbox{\rm d}\bar{t}$ in a slight
abuse of standard notation) which to the first order in the infinitesimal, $%
\varepsilon $, is
\begin{eqnarray}
\bar{A} &=&\int_{t_{0}}^{t_{1}}\left[ L+\varepsilon \left( \tau \displaystyle%
{\frac{\partial L}{\partial t}}+\eta \displaystyle{\frac{\partial L}{%
\partial q}}+\zeta \displaystyle{\frac{\partial L}{\partial \dot{q}}}+\dot{%
\tau}L\right) \right] \mbox{\rm d}t  \nonumber \\
&&+\varepsilon \left[ \tau {t_{1}}L(t_{1},q_{1},\dot{q}_{1})-\tau {t_{0}}%
L(t_{0},q_{0},\dot{q}_{0})\right] ,  \label{4425}
\end{eqnarray}%
where $\zeta =\dot{\eta}-\dot{q}\dot{\tau}$ and $L(t_{0},q_{0},\dot{q}_{0})$
and $L(t_{1},q_{1},\dot{q}_{1})$ are the values of $L$ at the endpoints $%
t_{0}$ and $t_{1}$ respectively.

We demonstrate the origin of the terms outside of the integral with the
upper limit. The lower limit is treated analogously.
\begin{eqnarray*}
\int^{\bar{t}_{1}} &=&\int^{t_{1}+\varepsilon \tau (t_{1})} \\
&=&\int^{t_{1}}+\int_{t_{1}}^{t_{1}+\varepsilon \tau (t_{1})} \\
&=&\varepsilon \int^{t_{1}}+\varepsilon \tau (t_{1})L(t_{1},q_{1},\dot{q}%
_{1})
\end{eqnarray*}%
to the first order in $\varepsilon $. We may rewrite (\ref{4425}) as
\[
\bar{A}=A+\varepsilon \int_{t_{0}}^{t_{1}}\left( \tau \displaystyle{\frac{%
\partial L}{\partial t}}+\eta \displaystyle{\frac{\partial L}{\partial q}}%
+\zeta \displaystyle{\frac{\partial L}{\partial \dot{q}}}+\dot{\tau}L\right) %
\mbox{\rm d}t+\varepsilon F,
\]%
where the number, $F$, is the value of the second term in crochets in (\ref%
{4425}). As $F$ depends only upon the endpoints, we may write it as
\[
F=-\int_{t_{0}}^{t_{1}}\dot{f}\mbox{\rm d}t,
\]%
where the sign is chosen as a matter of later convenience.\hfill

The generator, $\Gamma $, of the infinitesimal transformation, (\ref{4422}),
is a Noether symmetry of (\ref{4421}) if
\[
\bar{A}=A,
\]%
\textit{ie}
\[
\int_{t_{0}}^{t_{1}}\left( \tau \displaystyle{\frac{\partial L}{\partial t}}%
+\eta \displaystyle{\frac{\partial L}{\partial q}}+\zeta \displaystyle{\frac{%
\partial L}{\partial \dot{q}}}+\dot{\tau}L-\dot{f}\right) \mbox{\rm d}t=0%
\]%
from which it follows that
\begin{equation}
\dot{f}=\tau \displaystyle{\frac{\partial L}{\partial t}}+\eta \displaystyle{%
\frac{\partial L}{\partial q}}+\zeta \displaystyle{\frac{\partial L}{%
\partial \dot{q}}}+\dot{\tau}L.  \label{44210}
\end{equation}%
\textbf{Remark:} The symmetry is the generator of an infinitesimal
transformation which leaves the Action Integral invariant and the existence
of the symmetry has nothing to do with the Euler-Lagrange Equation of the
Calculus of Variations. The Euler-Lagrange equation follows from the
application of Hamilton's Principle in which $q$ is given a zero endpoint
variation. There is no such restriction on the infinitesimal transformations
introduced by Noether.\hfill

\section{Noether's Theorem}

We now invoke Hamilton's Principle for the Action Integral (\ref{4421}). We
observe that the zero-endpoint variation of (\ref{4421}) imposed by
Hamilton's Principle requires that (\ref{4421}) take a stationary value. Not
necessarily a minimum! The Principle of Least Action enunciated by Fermat in
1662 as `Nature always acts in the shortest ways' was raised to an even more
metaphysical status by Maupertuis \cite[p 254,p 267]{Dugas 88 a}. That the
principle applies in Classical (Newtonian) Mechanics is an accident of
metric! We can only wonder that the quasimystical principle has persisted
for over two centuries in what are supposed to be rational circles. In the
case of a first-order Lagrangian with a positive definite Hessian with
respect to $\dot{q}$ Hamilton's Principle gives a minimum. This is not
necessarily the case otherwise.\hfill

The Euler-Lagrange equation
\begin{equation}
\displaystyle{\frac{\partial L}{\partial q}}-\displaystyle{\frac{\mbox{\rm d}%
\,}{\mbox{\rm d}t}}\left( \displaystyle{\frac{\partial L}{\partial \dot{q}}}%
\right) =0  \label{4431}
\end{equation}%
follows from the application of Hamilton's Principle. We manipulate (\ref%
{44210}) as follows.
\begin{eqnarray*}
0 &=&\dot{f}-\tau \displaystyle{\frac{\partial L}{\partial t}}-\dot{\tau}%
L-\eta \displaystyle{\frac{\partial L}{\partial q}}-\left( \dot{\eta}-\dot{q}%
\dot{\tau}\right) \displaystyle{\frac{\partial L}{\partial \dot{q}}} \\
&=&\displaystyle{\frac{\mbox{\rm d}\,}{\mbox{\rm d}t}}\left( f-\tau L\right)
+\tau \left( \dot{q}\displaystyle{\frac{\partial L}{\partial q}}+\ddot{q}%
\displaystyle{\frac{\partial L}{\partial \dot{q}}}\right) +\dot{\tau}\left(
\dot{q}\displaystyle{\frac{\partial L}{\partial \dot{q}}}\right) \\
&&\qquad -\eta \displaystyle{\frac{\mbox{\rm d}\,}{\mbox{\rm d}t}}\left( %
\displaystyle{\frac{\partial L}{\partial \dot{q}}}\right) -\dot{\eta}%
\displaystyle{\frac{\partial L}{\partial \dot{q}}} \\
&=&\displaystyle{\frac{\mbox{\rm d}\,}{\mbox{\rm d}t}}\left[ f-\tau L-\left(
\eta -\tau \dot{q}\right) \displaystyle{\frac{\partial L}{\partial \dot{q}}}%
\right]
\end{eqnarray*}%
in the second line of which we have used the Euler-Lagrange Equation, (\ref%
{4431}), to change the coefficient of $\eta $. Hence we have a first
integral
\begin{equation}
I=f-\left[ \tau L+\left( \eta -\dot{q}\tau \right) \frac{\partial L}{%
\partial \dot{q}}\right]  \label{4433}
\end{equation}%
and an initial statement of Noether's Theorem.\hfill

\textbf{Noether's Theorem}: If the Action Integral of a first-order
Lagrangian, namely
\[
A=\int_{t_{0}}^{t_{1}}L\left( t,q,\dot{q}\right) \mbox{\rm d}t
\]%
is invariant under the infinitesimal transformation generated by the
differential operator
\[
\Gamma =\tau \partial _{t}+\eta _{i}\partial _{q_{i}},
\]%
there exists a function $f$ such that
\begin{equation}
\dot{f}=\tau \frac{\partial L}{\partial t}+\eta _{i}\frac{\partial L}{%
\partial q_{i}}+\zeta _{i}\frac{\partial L}{\partial \dot{q}_{i}}+\dot{\tau}%
L,  \label{4noe4}
\end{equation}%
where $\zeta _{i}=\dot{\eta}_{i}-\dot{q}_{i}\dot{\tau}$, and a first
integral given by
\[
I=f-\left[ \tau L+\left( \eta _{i}-\dot{q}_{i}\tau \right) \frac{\partial L}{%
\partial \dot{q}_{i}}\right] .\label{4noe6}
\]%
$\Gamma $ is called a Noether symmetry of $L$ and $I$ a Noetherian first
integral. The symmetry $\Gamma $ exists independently of the requirement
that the variation of the functional be zero. When the extra condition is
added, the first integral exists.\hfill

We note that there is not a one-to-one correspondence between a Noether
symmetry and a Noetherian integral. Once the symmetry is determined, the
integral follows with minimal effort. The converse is not so simple because,
given the Lagrangian and the integral, the symmetry is the solution of a
differential equation with an additional dependent variable, the function $f$
arising from the boundary terms. There can be an infinite number of
coefficient functions for a given first integral. The restriction of the
symmetry to a point symmetry may reduce the number of symmetries, too
effectively, to zero. 
The ease of determination of a Noetherian integral once the Noether symmetry
is known is in contrast to the situation for the determination of first
integrals in the case of Lie symmetries of differential equations. The
computation of the first integrals associated with a Lie symmetry can be a
highly nontrivial matter.\hfill

\section{Nonlocal Integrals}

We recall that the variable dependences of the coefficient functions $\tau $
and $\eta $ were not specified and \textit{do not enter into the derivation}
of the formulae\ for the coefficient functions or the first integral.
Consequently not only can we have the generalised symmetries of Noether's
paper but we can also have more general forms of symmetry such as nonlocal
symmetries \cite{Govinder 95 d, Pillay 96 a} without a single change in the
formalism. Of course, as has been noted for the calculation of first
integrals \cite{Leach 81 a} and symmetries in general \cite{Govinder 95 b},
the realities of computational complexity may force one to impose some
constraints on this generality. Once the Euler-Lagrange equation is invoked,
there is an automatic constraint on the degree of derivatives in any
generalised symmetry.\hfill

If one has a standard Lagrangian such as (\ref{4421}), a nonlocal Noether's
symmetry will usually produce a nonlocal integral through (\ref{4433}). In
that the total time derivative of this function is zero when the
Euler-Lagrange equation, (\ref{4431}), is taken into account, it is formally
a first integral. However, the utility of such a first integral is at best
questionable. \ Here Lie and Noether have generically differing outcomes. An
exponential nonlocal Lie symmetry can be expected to lead to a local first
integral whereas one could scarcely envisage the same for an exponential
nonlocal Noether symmetry.

On the other hand, if the Lagrangian was nonlocal, the combination of
nonlocal symmetry and nonlocal Lagrangian could lead to a local first
integral. However, we have not constructed a formalism to deal with nonlocal
Lagrangians -- as opposed to nonlocal symmetries -- and so we cannot simply
apply what we have developed above.

The introduction of a nonlocal term into the Lagrangian effectively
increases the order of the Lagrangian by one (in the case of a simple
integral) and the order of the associated Euler-Lagrange equation by two so
that for a Lagrangian regular in $\dot{q}$ instead of a second-order
differential equation we would have a fourth order differential equation in $%
q$. To avoid that the Lagrangian would have to be degenerate, \textit{ie}
linear, in $\dot{q}$ and this cannot, as is well-known, lead to a
second-order differential equation. It would appear that nonlocal symmetries
in the context of Noether's Theorem do not have the same potential as
nonlocal Lie symmetries of differential equations.\hfill

There is often some confusion of identity between Lie symmetries and Noether
symmetries. Although every Noether symmetry is a Lie symmetry of the
corresponding Euler-Lagrange equation, we stress that they have different
provenances. There is a difference which is more obvious in systems of
higher dimension. A Noether symmetry can only give rise to a single first
integral because of (\ref{4noe6}). In an $n$-dimensional system of
second-order ordinary differential equations a single Lie symmetry gives
rise to $(2n-1)$ first integrals \cite{leach01,leach02,leach03,leach04}%
.\hfill

\section{Extensions: one independent variable}

The derivation given above applies to a one-dimensional discrete system. The
theorem can be extended to continuous systems and systems of higher order.
The principle is the same. The mathematics becomes more complicated. We
simply quote the relevant results.\hfill

For a first-order Lagrangian with $n$ dependent variables
\begin{equation}
G=\tau \partial _{t}+\eta _{i}\partial _{q_{i}}  \label{4.1}
\end{equation}%
is a Noether symmetry of the Lagrangian, $L(t,q_{i},\dot{q}_{i})$, if there
exists a function $f$ such that
\begin{equation}
\dot{f}=\dot{\tau}L+\tau \frac{\partial L}{\partial t}+\eta _{i}\frac{%
\partial L}{\partial q_{i}}+\left( \dot{\eta}_{i}-\dot{q}_{i}\dot{\tau}%
\right) \frac{\partial L}{\partial \dot{q}_{i}}  \label{4.2}
\end{equation}%
and the corresponding Noetherian first integral is
\begin{equation}
I=f-\left[ \tau L+\left( \eta _{i}-\dot{q}_{i}\tau \right) \frac{\partial L}{%
\partial \dot{q}_{i}}\right]  \label{4.3}
\end{equation}%
which are the obvious generalisations of (\ref{44210}) and (\ref{4433})
respectively.\hfill

In the case of an $n$th-order Lagrangian in one dependent variable and one
independent variable, $L(t,q,\dot{q},\ldots,q^{(n)})$ with $q^{(n)} = %
\mbox{\rm d}^n q/\mbox{\rm d} t^n$, the Euler-Lagrange equation is
\begin{equation}
\sum_{j=0}^n (-1)^j\frac{\mbox{\rm d}^j\ }{\mbox{\rm d} t^j}\left (\frac{%
\partial L\ \ }{\partial q^{(j)}}\right).  \label{4.4}
\end{equation}
$\Gamma = \tau\partial_t + \eta\partial_q$ is a Noether symmetry if there
exists a function $f$ such that
\begin{equation}
\dot{f} = \dot{\tau}L + \tau \frac{\partial L}{\partial t} + \sum_{j=0}^n
(-1)^j\zeta^j \left (\frac{\partial L\ \ }{\partial q^{(j)}}\right),
\label{4.5}
\end{equation}
where
\begin{equation}
\zeta^j = \eta^{(j)} - \sum_{k=1}^j\left (
\begin{array}{c}
j \\
k%
\end{array}%
\right) q^{(j+1-k)}\tau^{(k)}.  \label{4.6}
\end{equation}
The expression for the first integral is
\begin{equation}
I = f - \left [ \tau L + \sum_{i=0}^{n-1}\sum_{j-0}^{n-1-i} (-1)^j\left
(\eta-\dot{q}\tau\right)^{(i)}\frac {\mbox{\rm d}^j\ }{\mbox{\rm d} t^j}%
\left(\frac {\partial L\phantom{(i+l)}}{\partial q^{(i+j+1)}}\right)\right ].
\label{4.7}
\end{equation}

\strut In the case of an $n$th-order Lagrangian in $m$ dependent variables
and one independent variable, $L(t,q_{k},\dot{q}_{k},\ldots ,q_{k}^{(n)})$
with $q_{k}^{(n)}=\mbox{\rm d}^{n}q_{k}/\mbox{\rm d}t^{n}$, $k=1,m$, the
Euler-Lagrange equation is
\begin{equation}
\sum_{j=0}^{n}(-1)^{j}\frac{\mbox{\rm d}^{j}\ }{\mbox{\rm d}t^{j}}\left(
\frac{\partial L\ \ }{\partial q_{k}^{(j)}}\right) ,\ k=1,m.  \label{4.8}
\end{equation}%
$\Gamma =\tau \partial _{t}+\sum_{k=1}^{m}\eta _{k}\partial _{q_{k}}$ is a
Noether symmetry if there exists a function $f$ such that
\begin{equation}
\dot{f}=\dot{\tau}L+\tau \frac{\partial L}{\partial t}+\sum_{k=1}^{m}%
\sum_{j=0}^{n}(-1)^{j}\zeta _{k}^{j}\left( \frac{\partial L\ \ }{\partial
q_{k}^{(j)}}\right) ,  \label{4.9}
\end{equation}%
where
\begin{equation}
\zeta _{k}^{j}=\eta _{k}^{(j)}-\sum_{k=1}^{m}\sum_{i=0}^{j}\left(
\begin{array}{c}
j \\
i%
\end{array}%
\right) q_{k}^{(j+1-i)}\tau ^{(i)}.  \label{4.10}
\end{equation}%
The expression for the first integral is
\begin{equation}
I=f-\left[ \tau
L+\sum_{k=1}^{m}\sum_{i=0}^{n-1}\sum_{j-0}^{n-1-i}(-1)^{j}\left( \eta _{k}-%
\dot{q_{k}}\tau \right) ^{(i)}\frac{\mbox{\rm d}^{j}\ }{\mbox{\rm d}t^{j}}%
\left( \frac{\partial L\phantom{(i+j+1)}}{\partial q_{k}^{(i+j+1)}}\right) %
\right] .  \label{4.11}
\end{equation}%
The expressions in (\ref{4.7}) and (\ref{4.11}), although complex enough,
conceal an even much greater complexity because each derivative with respect
to time is a total derivative and so affects all terms in the Lagrangian and
its partial derivatives.\hfill

\section{Observations}

In the case of a first-order Lagrangian with one independent variable it is
well-known \cite{Sarlet 81 a} that one can achieve a simplification in the
calculations of the Noether symmetry in the case that the Lagrangian has a
regular Hessian with respect to the $\dot{q}_{i}$. We suppose that we admit
generalised symmetries in which the maximum order of the derivatives present
in $\tau $ and the $\eta _{i}$ is one, \textit{ie} equal to the order of the
Lagrangian. Then the coefficient of each $\ddot{q}_{j}$ in (\ref{4.2}) is
separately zero since the Euler-Lagrange equation has not yet been invoked.
Thus we have
\begin{equation}
\frac{\partial f}{\partial \dot{q}_{j}}=\frac{\partial \tau }{\partial \dot{q%
}_{j}}L+\left( \frac{\partial \eta _{i}}{\partial \dot{q}_{j}}-\dot{q}_{i}%
\frac{\partial \tau }{\partial \dot{q}_{j}}\right) \frac{\partial L}{%
\partial \dot{q}_{i}}.  \label{4.13}
\end{equation}%
We differentiate (\ref{4.3}) with respect to $\dot{q}_{j}$ to obtain
\begin{equation}
\frac{\partial I}{\partial \dot{q}_{j}}=\frac{\partial f}{\partial \dot{q}%
_{j}}-\left[ \frac{\partial \tau }{\partial \dot{q}_{j}}L+\tau \frac{%
\partial L}{\partial \dot{q}_{j}}+\left( \frac{\partial \eta _{i}}{\partial
\dot{q}_{j}}-\delta _{ij}\tau -\dot{q}_{i}\frac{\partial \tau }{\partial
\dot{q}_{j}}\right) \frac{\partial L}{\partial \dot{q}_{i}}+\left( \eta _{i}-%
\dot{q}_{i}\tau \right) \frac{\partial ^{2}L}{\partial \dot{q}_{i}}{\partial
\dot{q}_{j}}\right] ,  \label{4.14}
\end{equation}%
where $\delta _{ij}$ is the usual Kronecker delta, which, when we take (\ref%
{4.13}) into account, gives
\begin{equation}
\frac{\partial I}{\partial \dot{q}_{j}}=-\left( \eta _{i}-\dot{q}_{i}\tau
\right) \frac{\partial ^{2}L}{\partial \dot{q}_{i}}{\partial \dot{q}_{j}}.
\label{4.15}
\end{equation}%
Consequently, if the Lagrangian is regular with respect to the $\dot{q}_{i}$%
, we have
\begin{equation}
\left( \eta _{i}-\dot{q}_{i}\tau \right) =-g_{ij}\frac{\partial I}{\partial
\dot{q}_{j}},  \label{4.16}
\end{equation}%
where
\[
g_{ik}\frac{\partial ^{2}L}{\partial \dot{q}_{k}}{\partial \dot{q}_{j}}%
=\delta _{ij}.
\]%
The relations (\ref{4.15}) and (\ref{4.16}) reveal two useful pieces of
information. The first is that the derivative dependence of the first
integral is determined by the nature of the generalised symmetry (modulo the
derivative dependence in the Lagrangian). The second is that there is a
certain freedom of choice in the structure of the functions $\tau $ and $%
\eta _{i}$ in the symmetry. Provided generalised symmetries are admitted,
there is no loss of generality in putting one of the coefficient functions
equal to zero. An attractive candidate is $\tau $ as it appears the most
frequently. The choice should be made before the derivative dependence of
the coefficient functions is assumed. We observe that in the case of a
`natural' Lagrangian, \textit{ie} one quadratic in the derivatives, the
first integrals can only be linear or quadratic in the derivatives if the
symmetry is assumed to be point.\hfill

\section{Examples}

\textbf{The free particle}

We consider the simple example of the free particle for which
\[
L=\mbox{$\frac{1}{2}$}y^{\prime 2}.
\]%
Equation (\ref{4noe4}) is
\begin{equation}
(\eta ^{\prime }-y^{\prime }\xi ^{\prime })y^{\prime }+\mbox{$\frac{1}{2}$}%
y^{\prime 2}=f^{\prime }.  \label{4noe7}
\end{equation}%
If we assume that $\Gamma $ is a Noether point symmetry, (\ref{4noe7}) gives
the following determining equations
\begin{eqnarray*}
y^{\prime 3}:\qquad \quad  &-\mbox{$\frac{1}{2}$}\displaystyle{\frac{%
\partial \xi }{\partial y}}=&0 \\
y^{\prime 2} &:&\displaystyle{\frac{\partial \eta }{y}}-\mbox{$\frac{1}{2}$}%
\displaystyle{\frac{\partial \xi }{\partial x}}=0 \\
y^{\prime 1} &:&\displaystyle{\frac{\partial \eta }{\partial x}}-%
\displaystyle{\frac{\partial f}{\partial y}}=0 \\
y^{\prime 0} &:&\displaystyle{\frac{\partial f}{\partial x}}=0
\end{eqnarray*}%
from which it is evident that
\begin{eqnarray*}
\xi  &=&a(x) \\
\eta  &=&\mbox{$\frac{1}{2}$}a^{\prime }y+b(x) \\
f &=&\mbox {$\frac {1} {4} $}a^{\prime \prime 2}+b^{\prime }y+c(x) \\
0 &=&\mbox {$\frac {1} {4} $}a^{\prime \prime \prime 2}+b^{\prime \prime
}y+c^{\prime }.
\end{eqnarray*}%
Hence
\begin{eqnarray*}
a &=&A_{0}+A_{1}x+A_{2}x^{2} \\
b &=&B_{0}+B_{1}x \\
c &=&C_{0}.
\end{eqnarray*}%
Because $c$ is simply an additive constant, it is ignored. There are five
Noether point symmetries which is the maximum for a one-dimensional system
\cite{Mahomed 93 a}. They and their associated first integrals are
\begin{eqnarray*}
\Gamma _{1}=\partial _{y} &\qquad \quad &I_{1}=-y^{\prime } \\
\Gamma _{2}=x\partial _{y} &&I_{2}=y-xy^{\prime } \\
\Gamma _{3}=\partial _{x} &&I_{3}=\mbox{$\frac{1}{2}$}y^{\prime 2} \\
\Gamma _{4}=x\partial _{x}+\mbox{$\frac{1}{2}$}y\partial _{y} &&I_{4}=-%
\mbox{$\frac{1}{2}$}y^{\prime }(y-xy^{\prime }) \\
\Gamma _{5}=x^{2}\partial _{x}+xy\partial _{y} &&I_{5}=\mbox{$\frac{1}{2}$}%
(y-xy^{\prime 2}.
\end{eqnarray*}%
The corresponding Lie algebra is isomorphic to $A_{5,40}$ \cite{Patera 77 a}%
. The algebra is structured as $2A_{1}\oplus _{s}sl(2,R)$ which is a proper
subalgebra of the Lie algebra for the differential equation for the free
particle, namely $sl(3,R)$ which is structured as $2A_{1}\oplus
_{s}\{sl(2,R)\oplus _{s}A_{1}\}\oplus _{s}2A_{1}$. The missing symmetries
are the homogeneity symmetry and the two noncartan symmetries. The absence
of the homogeneity symmetry emphasizes the distinction between the Lie and
Noether symmetries.\hfill

\textbf{Noether symmetries of a higher-order Lagrangian}

Suppose that $L = \mbox{$\frac{1}{2}$} y^{\prime \prime 2}$. The condition
for a Noether point symmetry is that
\begin{equation}
\zeta_2 \frac{\partial L}{\partial y^{\prime \prime }} + \xi^{\prime }L =
f^{\prime },  \label{4.20}
\end{equation}
where $\zeta_2 = \eta^{\prime \prime }- 2 y^{\prime \prime }\xi^{\prime }-
y^{\prime }\xi^{\prime \prime }$ so that (\ref{4.20}) becomes
\begin{equation}
(\eta^{\prime \prime }- 2 y^{\prime \prime }\xi^{\prime }- y^{\prime
}\xi^{\prime \prime }) y^{\prime \prime }+ \mbox{$\frac{1}{2}$} \xi^{\prime
}y^{\prime \prime 2 }= f^{\prime }.
\end{equation}
Assume a point transformation, \textit{ie} $\xi = \xi(x,y)$ and $\eta =
\eta(x,y)$. Then
\begin{eqnarray*}
&&\left[\frac{\partial ^ 2\eta}{\partial x ^ 2} + 2 y^{\prime }\frac {%
\partial ^ 2\eta}{\partial x}{\partial y} + y^{\prime 2 }\frac {\partial ^
2\eta}{\partial y ^ 2} + y^{\prime \prime }\frac {\partial\eta}{\partial y}
- 2 y^{\prime \prime }\left(\frac {\partial\xi}{\partial x} + y^{\prime }%
\frac {\partial\xi}{\partial y}\right)\right. \\
&&\mbox{} \left. - y^{\prime }\left(\frac {\partial ^ 2\xi}{\partial x ^ 2}
+ 2 y^{\prime }\frac {\partial ^ 2\xi}{\partial x}{\partial y} + y^{\prime 2
}\frac {\partial ^ 2 xi}{\partial y ^ 2} + y^{\prime \prime }\frac {%
\partial\xi}{\partial y}\right)\right] y^{\prime \prime }+ %
\mbox{$\frac{1}{2}$} \left(\frac {\partial\xi}{\partial x} + y^{\prime }%
\frac {\partial\xi}{\partial y}\right) y^{\prime \prime 2} \\
&=& \frac {\partial f}{\partial x} + y^{\prime }\frac {\partial f}{\partial y%
} + y^{\prime \prime }\frac {\partial f}{\partial y^{\prime }}.
\end{eqnarray*}
from the coefficient of $y^{\prime }y^{\prime \prime}{}^ 2$, \textit{%
videlicet}
\[
- \mbox {$\frac {5} {2} $} \frac {\partial\xi}{\partial y} = 0,
\]
we obtain
\[
\xi = a(x).
\]
The coefficient of $y^{\prime \prime}{}^ 2$,
\[
\frac{\partial\eta}{\partial y} - \mbox {$\frac {3} {2} $} \frac {\partial\xi%
}{\partial x} = 0,
\]
results in
\[
\eta = \mbox {$\frac {3} {2} $} a^{\prime }y + b(x)
\]
and the coefficient of $y^{\prime \prime }$,
\[
\frac{\partial ^ 2\eta}{\partial x ^ 2} + 2 y^{\prime }\frac {\partial ^
2\eta}{\partial x}{\partial y} + y^{\prime 2 }\frac {\partial ^ 2\eta}{%
\partial y ^ 2} - y^{\prime }\frac {\partial ^ 2\xi}{\partial x ^ 2} =
\frac {\partial f}{\partial y^{\prime }},
\]
gives $f$ as
\[
f = a^{\prime \prime }y^{\prime}{}^ 2+ (\mbox {$\frac {3} {2} $} a^{\prime
\prime \prime }y + b^{\prime \prime }) y^{\prime }+ c(x,y).
\]
The remaining terms give
\[
y^{\prime }\frac {\partial f}{\partial y} + \frac {\partial f}{\partial x} =
0,
\]
\textit{ie}
\[
y^{\prime }\left[\frac {3} {2} a^{\prime \prime \prime }y^{\prime }+ \frac {%
\partial c}{\partial y}\right] + a^{\prime \prime \prime }y^{\prime 2 }+
\left(\mbox {$\frac {3} {2} $}a^{iv}y + b^{\prime \prime \prime }\right)
y^{\prime }+ \displaystyle{\frac{\partial c}{\partial x}} = 0.
\]
The coefficient of $y^{\prime}{}^ 2$ is $\mbox {$\frac {5} {2} $} a^{\prime
\prime \prime }= 0 $ from which it follows that
\[
a = A_0 + A_1 x + A_2 x^2.
\]
The coefficient of $y^{\prime }$ is
\[
\frac {\partial c}{\partial y} + \mbox {$\frac {3} {2} $} a^{iv} y +
b^{\prime \prime \prime }= 0
\]
and so
\[
c = - b^{\prime \prime \prime }y + d(x).
\]
The remaining terms give
\[
\frac {\partial c}{\partial x} = 0,
\]
\textit{ie},
\[
- b^{iv} y + d^{\prime }= 0
\]
from which $d$ is a constant (and can therefore be ignored) and
\[
b = B_0 + B_1 x + B_2 x^2 + B_3 x^3.
\]
There are seven Noetherian point symmetries for $L = \mbox{$\frac{1}{2}$}
y^{\prime \prime 2}$. They and the associated `gauge functions' are
\[
\begin{array}{lll}
B_0: \qquad & \Gamma_1 = \partial_y & f_1 = 0 \\
B_1: & \Gamma_2 = x \partial_y & f_2 = 0 \\
B_2: & \Gamma_3 = x^2 \partial_y & f_3 = 2 xy^{\prime } \\
B_3: & \Gamma_4 = x^3 \partial_y & f_4 = 6 xy^{\prime }- 6y \\
A_0: & \Gamma_5 = \partial_x & f_5 = 0 \\
A_1: & \Gamma_6 = x \partial_x + \mbox {$\frac {3} {2} $} y \partial_y & f_6
= 0 \\
A_2: & \Gamma_7 = x^2 \partial_x + 3 xy \partial_y \qquad\quad & f_7 = 2
y^{\prime 2}.%
\end{array}
\]

\strut\hfill

The Euler-Lagrange equation for $L=\mbox{$\frac{1}{2}$}y^{\prime \prime}{}^ 2
$ is $y^{\left( iv\right) }=0$ which has Lie point symmetries the same as
the Noether point symmetries plus $\Gamma _{8}=y\partial _{y}$. Note that
there is a contrast here in comparison with the five Noether point
symmetries of $L=\mbox{$\frac{1}{2}$}y^{\prime}{}^ 2$ and the eight Lie
point symmetries of $y^{\prime \prime }=0$. The additional Lie symmetries
are $y\partial _{y}$ as above for $y^{iv}=0$ and the two noncartan
symmetries, $X_{1}=y\partial _{x}$ and $X_{2}=xy\partial _{x}+y^{2}\partial
_{y}$.\hfill

For $L=\mbox{$\frac{1}{2}$}y^{\prime \prime}{}^2$ the associated first
integrals have the structure
\[
I=f-\mbox{$\frac{1}{2}$}\xi y^{\prime \prime}{}^2+(\eta -y^{\prime }\xi
)y^{\prime \prime \prime }-(\eta ^{\prime }-y^{\prime \prime }\xi -y^{\prime
}\xi ^{\prime })y^{\prime \prime }
\]%
and are
\begin{eqnarray*}
I_{1} &=&y^{\prime \prime \prime } \\
I_{2} &=&xy^{\prime \prime \prime }-y^{\prime \prime } \\
I_{3} &=&x^{2}y^{\prime \prime \prime }-2xy^{\prime \prime }+2xy^{\prime } \\
I_{4} &=&x^{3}y^{\prime \prime \prime 2}y^{\prime \prime }+6xy^{\prime }-6y
\\
I_{5} &=&-y^{\prime }y^{\prime \prime \prime }+\mbox{$\frac{1}{2}$}y^{\prime
\prime 2} \\
I_{6} &=&-xy^{\prime }y+\mbox{$\frac{1}{2}$}xy^{\prime \prime 2}-%
\mbox{$\frac{1}{2}$}y^{\prime }y^{\prime \prime }+\mbox {$\frac {3} {2} $}%
yy^{\prime \prime \prime } \\
I_{7} &=&x(3y-xy^{\prime })y^{\prime \prime \prime }-(3y-xy^{\prime }-%
\mbox{$\frac{1}{2}$}x^{2}y^{\prime \prime })y^{\prime \prime
}+2y^{\prime}{}^ 2.
\end{eqnarray*}%
Note that $I_{1}$--$I_{4}$ associated with $\Gamma _{1}$--$\Gamma _{4}$
respectively are also integrals obtained by the Lie method. However, each
Noether symmetry produces just one first integral whereas each Lie symmetry
has three first integrals associated with it.\hfill

In this example only point Noether symmetries have been considered. One may
also determine symmetries which depend upon derivatives, effectively up to
the third order when one is calculating first integrals of the
Euler-Lagrange Equation.\hfill

\textbf{Omission of the gauge function}

In some statements of Noether's theorem the so-called gauge function, $f$,
is taken to be zero. In the derivation given here, $f$ comes from the
contribution of the boundary terms produced by the infinitesimal
transformation in $t$ and so is not a gauge function in the usual meaning of
the term. However, it does function as one since it is independent of the
trajectory in the extended configuration space and depends only upon the
evaluation of functions at the boundary (end points in a
one-degree-of-freedom case) and can conveniently be termed one especially in
the light of Boyer's Theorem \cite{Boyer 67 a}.

Consider the example $L=\mbox{$\frac{1}{2}$}y^{\prime}{}^ 2$ without $f$.
The equation for the symmetries,
\[
f^{\prime }=\xi \frac{\partial L}{\partial x}+\eta \frac{\partial L}{%
\partial y}+(\eta ^{\prime }-y^{\prime }\xi ^{\prime })\frac{\partial L}{%
\partial y^{\prime }}+\xi ^{\prime }L,
\]%
becomes
\[
0=\left( \frac{\partial \eta }{\partial x}+y^{\prime }\frac{\partial \eta }{%
\partial y}-y^{\prime }\frac{\partial \xi }{\partial x}-y^{\prime 2}\frac{%
\partial \xi }{\partial y}\right) y^{\prime }+\mbox{$\frac{1}{2}$}y^{\prime
2}\left( \frac{\partial \xi }{\partial x}+y^{\prime }\frac{\partial \xi }{%
\partial y}\right) .
\]%
We solve this in the normal way: The coefficients of $y^{\prime}{}^ 3$, $%
y^{\prime}{}^ 2$ and of $y^{\prime }$ give in turn
\begin{eqnarray*}
\xi &=&a(x) \\
\eta &=&\mbox{$\frac{1}{2}$}a^{\prime }y+b(x) \\
\mbox{$\frac{1}{2}$}a^{\prime \prime }y+b^{\prime } &=&0
\end{eqnarray*}%
which hold provided that
\[
a=A_{0}+A_{1}x\qquad \qquad b=B_{0},
\]%
\textit{ie} only three symmetries are obtained instead of the five when the
gauge function is present.\hfill

It makes no sense to omit the gauge function when the infinitesimal
transformation is restricted to be point and only in the dependent
variables.\hfill

\noindent\textbf{A higher-dimensional system}

We determine the Noether point symmetries and their associated first
integrals for
\[
L = \mbox{$\frac{1}{2}$} (\dot{x}^2 + \dot{y}^2)
\]
(which is the standard Lagrangian for the free particle in two dimensions).
The determining equation is
\begin{eqnarray*}
\frac {\partial f}{\partial t} + \dot{x} \frac {\partial f}{\partial x} +
\dot{y} \frac {\partial f}{\partial y} &=& \left( \frac {\partial\eta}{%
\partial t} + \dot{x}\frac {\partial\eta}{\partial x} + \dot{y} \frac {%
\partial\eta}{\partial y} - \dot{x} \left( \frac {\partial\xi}{\partial t} +
\dot{x} \frac {\partial\xi}{\partial x} + \dot{y}\frac {\partial\xi}{%
\partial y}\right)\right) \dot{x} \\
&&\mbox{}+ \left( \frac {\partial\zeta}{\partial t} + \dot{x} \frac {%
\partial\zeta}{\partial x} + \dot {y} \frac {\partial \zeta} {\partial y} -
\dot {y} \left( \frac {\partial\xi}{\partial t} + \dot { x} \frac {%
\partial\xi}{\partial x} + \dot { y} \frac {\partial\xi}{\partial y}%
\right)\right) \dot y,
\end{eqnarray*}
where $\eta_1 =\eta$ and $\eta_2 = \zeta$.

\strut\hfill

We separate by powers of $\dot{x}$ and $\dot{y}$. Firstly taking the
third-order terms we have
\begin{eqnarray*}
\dot{x}^{3}:\qquad -\displaystyle{\frac{\partial \xi }{\partial x}} &=&0 \\
\dot{x}^{2}\dot{y}:\qquad -\displaystyle{\frac{\partial \xi }{\partial y}}
&=&0 \\
\dot{x}\dot{y}^{2}:\quad -y\displaystyle{\frac{\partial \xi }{\partial x}}
&=&0 \\
\dot{y}^{3}:\qquad -\displaystyle{\frac{\partial \xi }{\partial y}} &=&0
\end{eqnarray*}%
which implies $\xi =a(t)$. We now consider the second-order terms: The
coefficient of $\dot{x}^{2}$ gives $\eta $ as $\eta =\dot{a}x+b(y,t)$, that
of $\dot{x}\dot{y}$ gives $\zeta $ as
\[
\zeta =-\frac{\partial b}{\partial y}x+c(y,t)
\]%
and that of $\dot{y}^{2}$ gives $c=\dot{a}y+d(t)$ and $b=e(t)y+g(t)$. Thus
far we have
\[
\xi =a(t)\qquad \eta =\dot{a}x+ey+g\qquad \zeta =-ex+\dot{a}y+d.
\]%
The coefficient of $\dot{x}$ gives $f$ as
\[
f=\mbox{$\frac{1}{2}$}\ddot{a}x^{2}+\dot{e}xy+\dot{g}x+K(y,t).
\]%
The coefficient of $\dot{y}$ requires that
\[
\dot{e}x+\frac{\partial K}{\partial y}=-\dot{e}x+\ddot{a}y+\dot{d}
\]%
which implies
\[
\dot{e}=0\qquad \quad K=\mbox{$\frac{1}{2}$}\ddot{a}y^{2}+\dot{d}y+h(t).
\]%
The remaining term requires that
\[
\mbox{$\frac{1}{2}$}\mathinner{\buildrel\vbox{\kern5pt\hbox{...}}\over{a}}%
x^{2}+\ddot{g}x+\mbox{$\frac{1}{2}$}\mathinner{\buildrel\vbox{\kern5pt%
\hbox{...}}\over{a}}y^{2}+\ddot{d}y+\dot{h}=0
\]%
whence
\begin{eqnarray*}
a &=&A_{0}+A_{1}t+A_{2}t^{2} \\
g &=&G_{0}+G_{1}t \\
d &=&D_{0}+D_{1}t \\
h &=&H_{0}
\end{eqnarray*}%
(we ignore $H_0$ as it is an additive constant to $f$).\hfill

The coefficient functions are
\begin{eqnarray*}
\xi &=&A_{0}+A_{1}t+A_{2}t^{2} \\
\eta &=&(A_{1}+2A_{2}t)x+E_{0}y+G_{0}+G_{1}t \\
\zeta &=&-E_{0}x+(A_{1}+2A_{2}t)y+D_{0}+D_{1}t
\end{eqnarray*}%
and the gauge function is
\[
f=A_{2}x^{2}+G_{1}x+A_{2}y^{2}+D_{1}y.
\]%
We obtain three symmetries from $a$, namely
\begin{eqnarray*}
\Gamma _{1} &=&\partial _{t} \\
\Gamma _{2} &=&t\partial _{t}+x\partial _{x}+y\partial _{y} \\
\Gamma _{3} &=&t^{2}\partial _{t}+2t\left( x\partial _{x}+y\partial
_{y}\right)
\end{eqnarray*}%
which form $sl(2,R)$, one from $e$,
\[
\Gamma _{4}=y\partial _{x}-x\partial _{y}
\]%
which is $so(2)$, and four from $g$ and $d$, namely
\begin{eqnarray*}
\Gamma _{5} &=&\partial _{x} \\
\Gamma _{6} &=&t\partial _{x} \\
\Gamma _{7} &=&\partial _{y} \\
\Gamma _{8} &=&t\partial _{y}.
\end{eqnarray*}%
The last four are the `solution' symmetries and form the Lie algebra $4A_{1}$%
.\hfill

\section{More than one independent variable: preliminaries}

\subsection{Euler-Lagrange equation}

Noether's original formulation of her theorem was in the context of
Lagrangians for functions of several independent variables. We have
deliberately separated the case of one independent variable from the general
discussion to be able to present the essential ideas in as simple a form as
possible. The discussion of the case of several independent variables is
inherently more complex simply from a notational point of view although
there is no real increase in conceptual difficulty.
\hfill

We commence with the simplest instance of a Lagrangian of this class which
is $L(t,x,u,u_{t},u_{x})$, \textit{ie} one dependent variable, $u$, and two
dependent variables, $t$ and $x$. We recall the derivation of the
Euler-Lagrange equation for $u(t,x)$ consequent upon the application of
Hamilton's Principle. In the Action Integral
\begin{equation}
A=\int_{\Omega }L\left( t,x,u,u_{t},u_{x}\right) \mbox{\rm d}x\mbox{\rm d}t
\label{4.40}
\end{equation}%
we introduce an infinitesimal variation of the dependent variable,
\begin{equation}
\bar{u}=u+\varepsilon v(t,x),
\end{equation}%
where $\varepsilon $ is the infinitesimal parameter, $v(t,x)$ is
continuously differentiable in both independent variables and is required to
be zero on the boundary, $\partial \Omega $, of the domain of integration, $%
\Omega $, which in this introductory case is some region in the $(t,x)$
plane. Otherwise $v$ is an arbitrary function. We have
\begin{equation}
\bar{A}=\int_{\Omega }L\left( t,x,\bar{u},\bar{u}_{t},\bar{u}_{x}\right) %
\mbox{\rm d}x\mbox{\rm d}t  \label{4.41}
\end{equation}%
and we require the Action Integral to take a stationary value, \textit{ie} $%
\delta A=\bar{A}-A$ be zero. Now
\begin{eqnarray}
\delta A &=&\int_{\Omega }\left[ L\left( t,x,\bar{u},\bar{u}_{t},\bar{u}%
_{x}\right) -L\left( t,x,u,u_{t},u_{x}\right) \right] \mbox{\rm d}x%
\mbox{\rm
d}t  \nonumber \\
&=&\varepsilon \int_{\Omega }\left[ \displaystyle{\frac{\partial L}{\partial
u}}v+\displaystyle{\frac{\partial L}{\partial {u}_{t}}}\displaystyle{\frac{v%
}{t}}+\displaystyle{\frac{L}{{u}_{x}}}\displaystyle{\frac{v}{x}}\right] %
\mbox{\rm d}x\mbox{\rm d}t+O(\varepsilon ^{2})  \nonumber \\
&=&\varepsilon \int_{\Omega }\left[ \displaystyle{\frac{L}{u}}-\displaystyle{%
\frac{\partial \,}{\partial t}}\left( \displaystyle{\frac{L}{{u}_{t}}}%
\right) -\displaystyle{\frac{\partial \,}{\partial x}}\left( \displaystyle{%
\frac{\partial L}{\partial \bar{u}_{x}}}\right) \right] v\mbox{\rm d}t%
\mbox{\rm d}x  \nonumber \\
&&\quad +\int_{\partial \Omega }\left[ \displaystyle{\frac{\partial L}{%
\partial u_{t}}}\mbox{\rm d}t+\displaystyle{\frac{\partial L}{\partial {u}%
_{x}}}\mbox{\rm d}x\right] v+O(\varepsilon ^{2}).  \label{4.42}
\end{eqnarray}%
The second integral in (\ref{4.42}) is the sum of four integrals, two along
each of the intervals $(t_{1},t_{2})$ and $(x_{1},x_{2})$ with $x=x_{1}$ and
$x=x_{2}$ for the two integrals with respect to $t$ and $t=t_{1}$ and $%
t=t_{2}$ for the two integrals with respect to $x$. Because $v$ is zero on
the boundary, this term must be zero. As $v$ is otherwise arbitrary, the
expression within the crochets in the first integral must be zero and so we
have the Euler-Lagrange equation
\begin{equation}
\displaystyle{\frac{\partial L}{\partial u}}-\displaystyle{\frac{\partial \,%
}{\partial t}}\left( \displaystyle{\frac{\partial L}{\partial {u}_{t}}}%
\right) -\displaystyle{\frac{\partial \,}{\partial x}}\left( \displaystyle{%
\frac{\partial L}{\partial \bar{u}_{x}}}\right) =0.  \label{4.43}
\end{equation}%
A \textit{conservation law} for (\ref{4.43}) is a vector-valued function, $%
\mathbf{f}$, of $t,x,u$ and the partial derivatives of $u$ which is
divergence free, \textit{ie}
\begin{eqnarray}
\mbox{\rm div.}\mathbf{f} &=&\displaystyle{\frac{\partial f^{1}}{\partial t}}%
+\displaystyle{\frac{\partial f^{2}}{\partial x}}  \nonumber \\
&=&0.  \label{4.44}
\end{eqnarray}%
In (\ref{4.44}) the operators $\partial _{t}$ and $\partial _{x}$ are
operators of total differentiation with respect to $t$ and $x$,
respectively, and henceforth we denote these operators by $D_{t}$ and $D_{x}$%
. The standard symbol for partial differentiation, $\partial _{A}$,
indicates differentiation solely with respect to $A$. In this notation (\ref%
{4.43}) and (\ref{4.44}) become respectively
\begin{eqnarray}
\displaystyle{\frac{\partial L}{\partial u}}-D_{t}\displaystyle{\frac{%
\partial L}{\partial u_{t}}}-D_{x}\displaystyle{\frac{\partial L}{\partial
u_{x}}} &=&0\quad \mbox{\rm and}  \label{4.45} \\
\mbox{\rm div.}\mathbf{f}=D_{t}f^{1}+D_{x}f^{2} &=&0.  \label{4.46}
\end{eqnarray}%
Naturally there is no distinction between $D_{t}$, $\partial u/\partial t$
and $u_{t}$,
likewise for the derivatives with respect to $x$.\hfill

\subsection{Noether's Theorem for $L(t,x,u,u_t,u_x)$}

We introduce into the Action Integral an infinitesimal transformation,
\begin{equation}
\bar{t}=t+\varepsilon \tau \quad \bar{x}=x+\varepsilon \xi \quad \bar{u}%
=u+\varepsilon \eta \,  \label{4.46a}
\end{equation}%
generated by the differential operator
\begin{equation}
\Gamma =\tau \partial _{t}+\xi \partial _{x}+\eta \partial _{u},
\label{4.47}
\end{equation}%
which, because the Lagrangian depends upon $u_{t}$ and $u_{x}$, we extend
once to give
\begin{equation}
\Gamma ^{\lbrack 1]}=\Gamma +\left( D_{t}\eta -u_{t}D_{t}\tau -u_{t}D_{t}\xi
\right) \partial _{u_{t}}+\left( D_{x}\eta -u_{x}D_{t}\tau -u_{x}D_{x}\xi
\right) \partial _{u_{x}}.  \label{4.48}
\end{equation}%
The coefficient functions $\tau $, $\xi $ and $\eta $ may depend upon
derivatives of $u$ as well as $t$, $x$ and $u$.\hfill

The change in the Action due to the infinitesimal transformation is given by
\begin{eqnarray}
\delta A &=&\bar{A}-A  \nonumber \\
&=&\int_{\bar{\Omega}}L\left( \bar{t},\bar{x},\bar{u},\bar{u}_{\bar{t}},\bar{%
u}_{\bar{x}}\mbox{\rm d}\bar{t}\right) \mbox{\rm d}\bar{x}-\int_{{\Omega }%
}L\left( {t},{x},{u},{u}_{t},{u}_{{x}}\right) \mbox{\rm d}t\mbox{\rm d}x,
\label{4.49}
\end{eqnarray}%
where $\bar{\Omega}$ is the transformed domain. We recall that Noether's
Theorem comes in two parts. In the first part, with which we presently deal,
the discussion is about the Action Integral and not the Variational
Principle. Consequently there is no reason why the domain over which
integration takes place should be the same before and after the
transformation. Equally there is no reason to require that the coefficient
functions vanish on the boundary of $\Omega $. To make progress in the
analysis of (\ref{4.49}) we must reconcile the variables and domains of
integration. For the variables of integration we have
\begin{eqnarray}
\mbox{\rm d}\bar{t}\mbox{\rm d}\bar{x} &=&\frac{\partial \left( \bar{t},\bar{%
x}\right) }{\partial \left( t,x\right) }\mbox{\rm d}t\mbox{\rm d}x  \nonumber
\\
&=&\left\vert
\begin{array}{rr}
D_{t}\bar{t} & D_{t}\bar{x} \\
D_{x}\bar{t} & D_{x}\bar{x}%
\end{array}%
\right\vert \mbox{\rm d}t\mbox{\rm d}x  \nonumber \\
&=&\left\vert
\begin{array}{rr}
1+\varepsilon D_{t}\tau & \varepsilon D_{t}\xi \\
\varepsilon D_{x}\tau & 1+\varepsilon D_{x}\xi%
\end{array}%
\right\vert \mbox{\rm d}t\mbox{\rm d}x  \nonumber \\
&=&\left[ 1+\varepsilon \left( D_{t}\tau +D_{x}\xi \right) +O\left(
\varepsilon ^{2}\right) \right] \mbox{\rm d}t\mbox{\rm d}x.  \label{4.50}
\end{eqnarray}%
For the domain we have simply that
\begin{equation}
\bar{\Omega}=\Omega +\delta \Omega
\end{equation}%
which, as the transformation is infinitesimal, in general means the
evaluation of the surface integral and in this two-dimensional case the
evaluation of the line integral along the boundary of the original domain.
Although this domain is arbitrary, it is fixed for the Variational Principle
we are using. We can use the Divergence Theorem to express this in terms of
the volume integral over the original domain of the divergence of some
vector-valued function. Combining these considerations with (\ref{4.49}) and
(\ref{4.50}) and expanding the integrand of the first integral in (\ref{4.49}%
) as a Taylor series we can write the condition that the Action Integral be
invariant under the infinitesimal transformation as
\begin{eqnarray}
0 &=&\int_{\Omega }\left\{ L+\varepsilon \left[ \tau \displaystyle{\frac{%
\partial L}{\partial t}}+\xi \displaystyle{\frac{\partial L}{\partial x}}%
+\eta \displaystyle{\frac{\partial L}{\partial u}}+\left( D_{t}\eta
-u_{t}D_{t}\tau -u_{t}D_{t}\xi \right) \displaystyle{\frac{\partial L}{%
\partial u_{t}}}\right. \right.  \nonumber \\
&&\left. \left. +\left( D_{x}\eta -u_{x}D_{t}\tau -u_{x}D_{x}\xi \right) %
\displaystyle{\frac{\partial L}{\partial u_{x}}}\right] -\varepsilon %
\mbox{\rm div.}\mathbf{F}\right\} \left[ 1+\varepsilon \left( D_{t}\tau
+D_{x}\xi \right) \right] \mbox{\rm d}t\mbox{\rm d}x  \nonumber \\
&&-\int_{\Omega }L\mbox{\rm d}t\mbox{\rm d}x+O\left( \varepsilon ^{2}\right)
,  \label{4.51}
\end{eqnarray}%
where $\mathbf{F}$ represents the contribution from the boundary term. If we
require that this be true for any domain in which the Lagrangian is validly
defined, the first-order term in (\ref{4.51}) gives the condition for the
Lagrangian to possess a Noether symmetry, \textit{videlicet}
\begin{eqnarray}
\mbox{\rm div.}\mathbf{F} &=&\left( D_{t}\tau +D_{x}\xi \right) L+\tau %
\displaystyle{\frac{\partial L}{\partial t}}+\xi \displaystyle{\frac{%
\partial L}{\partial x}}+\eta \displaystyle{\frac{\partial L}{\partial u}}
\nonumber \\
&&+\left( D_{t}\eta -u_{t}D_{t}\tau -u_{x}D_{t}\xi \right) \displaystyle{%
\frac{\partial L}{\partial u_{t}}}+\left( D_{x}\eta -u_{t}D_{x}\tau
-u_{x}D_{x}\xi \right) \displaystyle{\frac{\partial L}{\partial u_{x}}}.
\label{4.52}
\end{eqnarray}%
The rest is just a matter of computation! There does not appear to be code
which enables one to solve (\ref{4.52}) for a given Lagrangian even for
point symmetries. One is advised \cite[p 273]{Bluman 89 a} to calculate the
(generalised) symmetries of the corresponding Euler-Lagrange equation and
then test whether there exists an $\mathbf{F}$ such that each of these
symmetries in turn satisfies (\ref{4.52}).

There is a theorem in Olver \cite[p 326]{Olver 86 a} that the set of
generalised symmetries of the Euler-Lagrange equation contains the set of
generalised Noether symmetries of the Lagrangian. A purist could well prefer
to be able to solve (\ref{4.52}) directly. Alan Head, the distinguished
Australian scientist, who wrote one of the more successful codes for
differential equations in 1978, considered the effort involved to write the
requisite code twenty years later excessive when the indirect route was
available (A K Head, private communication, December, 1997).
\hfill

A conservation law corresponding to a Noether symmetry `derived' from (\ref%
{4.52}) is obtained when the Euler-Lagrange Equation is taken into account.
We rewrite the right side of (\ref{4.52}) and have
\begin{eqnarray}
\mbox{\rm div.}\mathbf{F} &=&D_{t}\left[ \tau L+\eta \displaystyle{\frac{%
\partial L}{\partial u_{t}}}\right] +D_{x}\left[ \xi L+\eta \displaystyle{%
\frac{\partial L}{\partial u_{x}}}\right] +\tau \displaystyle{\frac{\partial
L}{\partial t}}+\xi \displaystyle{\frac{\partial L}{\partial x}}+\eta %
\displaystyle{\frac{\partial L}{\partial u}}  \nonumber \\
&&\quad -\left( u_{t}D_{t}\tau +u_{x}D_{t}\xi \right) \displaystyle{\frac{%
\partial L}{\partial u_{t}}}-\left( u_{t}D_{x}\tau +u_{x}D_{x}\xi \right) %
\displaystyle{\frac{\partial L}{\partial u_{x}}}  \nonumber \\
&&\quad -\tau D_{t}L-\xi D_{x}L-\eta D_{t}\displaystyle{\frac{\partial L}{%
\partial u_{t}}}-\eta D_{x}\displaystyle{\frac{\partial L}{\partial u_{x}}}
\nonumber \\
&=&D_{t}\left[ \tau L+\eta \displaystyle{\frac{\partial L}{\partial u_{t}}}%
\right] +D_{x}\left[ \xi L+\eta \displaystyle{\frac{\partial L}{\partial
u_{x}}}\right] +\tau \displaystyle{\frac{\partial L}{\partial t}}+\xi %
\displaystyle{\frac{\partial L}{\partial x}}-\left( u_{t}D_{t}\tau
+u_{x}D_{t}\xi \right) \displaystyle{\frac{\partial L}{\partial u_{t}}}
\nonumber \\
&&\quad -\left( u_{t}D_{x}\tau +u_{x}D_{x}\xi \right) \displaystyle{\frac{%
\partial L}{\partial u_{x}}}-\tau D_{t}L-\xi D_{x}L  \nonumber \\
&=&D_{t}\left[ \tau L+\left( \eta -u_{t}\tau -u_{x}\xi \right) \displaystyle{%
\frac{\partial L}{\partial u_{t}}}\right] +D_{x}\left[ \xi L+\left( \eta
-u_{t}\tau -u_{x}\xi \right) \displaystyle{\frac{\partial L}{\partial u_{x}}}%
\right]  \nonumber \\
&&\quad +\tau \left[ \displaystyle{\frac{\partial L}{\partial t}}%
-D_{t}L+D_{t}\left( u_{t}\displaystyle{\frac{\partial L}{\partial u_{t}}}%
\right) +D_{x}\left( u_{t}\displaystyle{\frac{\partial L}{\partial u_{x}}}%
\right) \right]  \nonumber \\
&&+\xi \left[ \displaystyle{\frac{\partial L}{\partial x}}%
-D_{x}L+D_{t}\left( u_{x}\displaystyle{\frac{\partial L}{\partial u_{t}}}%
\right) +D_{x}\left( u_{x}\displaystyle{\frac{\partial L}{\partial u_{x}}}%
\right) \right]  \nonumber \\
&=&D_{t}\left[ \tau L+\left( \eta -u_{t}\tau -u_{x}\xi \right) \displaystyle{%
\frac{\partial L}{\partial u_{t}}}\right] +D_{x}\left[ \xi L+\left( \eta
-u_{t}\tau -u_{x}\xi \right) \displaystyle{\frac{\partial L}{\partial u_{x}}}%
\right]  \label{4.53}
\end{eqnarray}%
when the Euler-Lagrange Equation is taken into account. Hence there is the
vector of the conservation law
\begin{equation}
\mathbf{I}=\mathbf{F}-\left[ \tau L+\left( \eta -u_{t}\tau -u_{x}\xi \right) %
\displaystyle{\frac{\partial L}{\partial u_{t}}}\right] \mathbf{e}_{t}-\left[
\xi L+\left( \eta -u_{t}\tau -u_{x}\xi \right) \displaystyle{\frac{\partial L%
}{\partial u_{x}}}\right] \mathbf{e}_{x},  \label{4.54}
\end{equation}%
where $\mathbf{e}_{t}$ and $\mathbf{e}_{x}$ are the unit vectors in the $%
(t,x)$ plane.\hfill

We consider the simple example of the Lagrangian
\[
L=\mbox {$\frac {1} {12} $}\left( u_{x}\right) ^{4}+\mbox{$\frac{1}{2}$}%
\left( u_{t}\right) ^{2}.
\]%
The condition for the existence of a Noether symmetry, (\ref{4.52}), becomes
\begin{eqnarray}
\mbox{\rm div.}\mathbf{F} &=&\left( D_{t}\tau +D_{x}\xi \right) \left(
\mbox
{$\frac {1} {12} $}\left( u_{x}\right) ^{4}+\mbox{$\frac{1}{2}$}\left(
u_{t}\right) ^{2}\right) +\left( D_{t}\eta -u_{t}D_{t}\tau -u_{x}D_{t}\xi
\right) u_{t}  \nonumber \\
&&\quad +\left( D_{x}\eta -u_{t}D_{x}\tau -u_{x}D_{x}\xi \right)
\mbox
{$\frac {1} {3} $}u_{x}^{3}.  \label{4.55}
\end{eqnarray}%
The Lagrangian has the Euler-Lagrange Equation
\begin{equation}
u_{x}^{2}u_{xx}+u_{tt}=0.  \label{4.56}
\end{equation}%
The Lie point symmetries of (\ref{4.56} are
\begin{equation}
\begin{array}{ll}
\Gamma _{1}=\partial _{t} & \Gamma _{4}=t\partial _{u} \\
\Gamma _{2}=\partial _{x} & \Gamma _{5}=t\partial _{t}-u\partial _{u} \\
\Gamma _{3}=\partial _{u} & \Gamma _{6}=x\partial _{x}+2u\partial _{u}.%
\end{array}
\label{4.57}
\end{equation}%
The Lie point symmetries $\Gamma _{1}$--$\Gamma _{4}$ give a zero vector $%
\mathbf{F}$ except for $\Gamma _{4}$ which gives $(u,0)$. The symmetries $%
\Gamma _{5}$ and $\Gamma _{6}$ give nonlocal vectors and so nonlocal
conservation laws, which could be interpreted as meaning that they are not
Noether symmetries for the given Lagrangian. The four local conservation
laws are
\begin{eqnarray*}
\mathbf{I}_{1} &=&\left( \mbox{$\frac{1}{2}$}u_{t}^{2}-u_{x}^{4},%
\mbox
{$\frac {1} {12} $}u_{t}u_{x}^{3}\right) \\
\mathbf{I}_{2} &=&\left( u_{t}u_{x},\mbox {$\frac {1} {4} $}u_{x}^{4}-%
\mbox{$\frac{1}{2}$}u_{t}^{2}\right) \\
\mathbf{I}_{3} &=&\left( u_{t},\mbox {$\frac {1} {3} $}u_{x}^{3}\right) \\
\mathbf{I}_{4} &=&\left( u-tu_{t},-\mbox {$\frac {1} {3} $}tu_{x}^{3}\right)
.
\end{eqnarray*}%
\hfill

\section{The general Euler-Lagrange Equation}

In the case of a $p$th-order Lagrangian in $m$ dependent variables, $u_{i}$,
$i=1,m$, and $n$ independent variables, $x_{j}$, $j=1,n$, the Lagrangian, $%
L(x,u,u_{1},\ldots ,u_{p})$, under an infinitesimal transformation
\[
\bar{u}^{i}(x)=u^{i}(x)+\varepsilon v^{i}(x),
\]%
where $\varepsilon $ is the parameter of smallness and $v(x)$ is $k-1$ times
differentiable and zero on the boundary $\partial \Omega $ of the domain of
integration $\Omega $ of the Action Integral,
\begin{equation}
A=\int_{\Omega }L\left( x,u,u_{1},\ldots ,u_{p}\right) \mbox{\rm d}x,
\end{equation}%
becomes
\begin{eqnarray}
\bar{L} &=&L\left( \bar{x},\bar{u},\bar{u}_{1},\ldots ,\bar{u}_{p}\right)
\nonumber \\
&=&L\left( x,u,u_{1},\ldots ,u_{p}\right) +\varepsilon v_{j_{1},j_{2},\ldots
,j_{k}}^{i}\displaystyle{\frac{\partial L}{\partial u_{j_{1},j_{2},\ldots
,j_{k}}^{i}}}+O\left( \varepsilon ^{2}\right)
\end{eqnarray}%
in which summation over repeated indices is implied and $i=1,m$, $j=1,n$ and
$k=0,p$. The variation in the Action Integral is
\begin{eqnarray}
\delta A &=&\int_{\Omega }\left[ \bar{L}-L\right] \mbox{\rm d}x  \nonumber \\
&=&\varepsilon \int_{\Omega }v_{j_{1},j_{2},\ldots ,j_{k}}^{i}\displaystyle{%
\frac{\partial L}{\partial u_{j_{1},j_{2},\ldots ,j_{k}}^{i}}}\mbox{\rm d}%
x+O\left( \varepsilon ^{2}\right) .  \label{4.58}
\end{eqnarray}%
We consider one set of terms in (\ref{4.58}) with summation only over $j_{k}$%
.
\begin{eqnarray}
&&\int_{\Omega }v_{j_{1},j_{2},\ldots ,j_{k}}^{i}\displaystyle{\frac{%
\partial L}{\partial u_{j_{1},j_{2},\ldots ,j_{k}}^{i}}}\mbox{\rm d}x
\nonumber \\
&=&\int_{\Omega }\left\{ D_{j_{k}}\left[ v_{j_{1},j_{2},\ldots ,j_{k-1}}^{i}%
\displaystyle{\frac{\partial L}{\partial u_{j_{1},j_{2},\ldots ,j_{k}}^{i}}}%
\right] -v_{j_{1},j_{2},\ldots ,j_{k-1}}^{i}D_{j_{k}}\left[ \displaystyle{%
\frac{\partial L}{\partial u_{j_{1},j_{2},\ldots ,j_{k}}^{i}}}\right]
\right\} \mbox{\rm d}x  \nonumber \\
&=&\int_{\partial \Omega }D_{j_{k}}\left[ v_{j_{1},j_{2},\ldots ,j_{k-1}}^{i}%
\displaystyle{\frac{\partial L}{\partial u_{j_{1},j_{2},\ldots ,j_{k}}^{i}}}%
\right] n_{j_{k}}\mbox{\rm d}\sigma -\int_{\Omega }v_{j_{1},j_{2},\ldots
,j_{k-1}}^{i}D_{j_{k}}\left[ \displaystyle{\frac{\partial L}{\partial
u_{j_{1},j_{2},\ldots ,j_{k}}^{i}}}\right] \mbox{\rm d}x  \nonumber \\
&=&\int_{\Omega }v_{j_{1},j_{2},\ldots ,j_{k-1}}^{i}D_{j_{k}}\left[ %
\displaystyle{\frac{\partial L}{\partial u_{j_{1},j_{2},\ldots ,j_{k}}^{i}}}%
\right] \mbox{\rm d}x,  \label{4.60}
\end{eqnarray}%
where on the right side in passing from the first line to the second line we
have made use of the Divergence Theorem and from the second to the third the
requirement that $v$ and its derivatives up to the $(p-1)th$ be zero on the
boundary. If we apply this stratagem repeatedly to (\ref{4.60}), we
eventually obtain that
\begin{equation}
\int_{\Omega }v_{j_{1},j_{2},\ldots ,j_{k}}^{i}\displaystyle{\frac{\partial L%
}{\partial u_{j_{1},j_{2},\ldots ,j_{k}}^{i}}}\mbox{\rm d}%
x=(-1)^{k}\int_{\Omega }v^{i}D_{j_{1}}D_{j_{2}}\ldots D_{j_{k}}\displaystyle{%
\frac{\partial L}{\partial u_{j_{1},j_{2},\ldots ,j_{k}}^{i}}}\mbox{\rm d}x.
\label{4.61}
\end{equation}

\strut\hfill

We substitute (\ref{4.61}) into (\ref{4.58}) to give
\begin{equation}
\delta A=\varepsilon (-1)^{k}\int_{\Omega }v^{i}D_{j_{1}}D_{j_{2}}\ldots
D_{j_{k}}\displaystyle{\frac{\partial L}{\partial u_{j_{1},j_{2},\ldots
,j_{k})}^{i}}}\mbox{\rm d}x+O\left( \varepsilon ^{2}\right) .  \label{4.62}
\end{equation}%
Hamilton's Principle requires that $\delta A$ be zero for a zero-boundary
variation. As the functions $v^{i}(x)$ are arbitrary subject to the
differentiability condition, the integrand in (\ref{4.62}) must be zero for
each value of the index $i$ and so we obtain the $m$ Euler-Lagrange
Equations
\begin{equation}
(-1)^{k}D_{j_{1}}D_{j_{2}}\ldots D_{j_{k}}\displaystyle{\frac{\partial L}{%
\partial u_{j_{1},j_{2},\ldots ,j_{k}}^{i}}}=0,\quad i=1,m,  \label{4.63}
\end{equation}%
with the summation on $j$ being from one to $n$ and on $k$ from zero to $p$%
.\hfill

\section{Noether's Theorem: original formulation}

Under the infinitesimal transformation
\begin{equation}
\bar{x}^j = x^j + \varepsilon\xi^j \qquad \bar{u}^i = u^i + \varepsilon\eta^i
\label{4.64}
\end{equation}
of both independent and dependent variables generated by the differential
operator
\begin{equation}
\Gamma = \xi_j\partial_{x_j} + \eta_i\partial_{u_i},  \label{4.12}
\end{equation}
in which summation on $i$ and $j$ from $1$ to $m$ and from $1$ to $n$
respectively is again implied, the Action Integral,
\begin{equation}
A = \int_{\Omega}L\left(x,u,u_1,\ldots,u_p\right)\mbox{\rm d} x,
\end{equation}
becomes
\begin{eqnarray}
\bar{A} & = & \int_{\bar{\Omega}}L\left(\bar{x},\bar{u},\bar{u}_1,\ldots,%
\bar{u}_p\right)\mbox{\rm d}\bar{x}  \nonumber \\
& = &
\int_{\Omega+\delta\Omega}L\left(x+\varepsilon\xi,u+\varepsilon\eta,u_1+%
\varepsilon\eta^{(1)}, \ldots,u_p+\varepsilon\eta^{(p)}\right)J\left(\bar{x}%
,x\right)\mbox{\rm d} x.  \label{4.65}
\end{eqnarray}
The notation $\delta\Omega$ indicates the infinitesimal change in the domain
of integration $\Omega$ induced by the infinitesimal transformation of the
independent variables.

\strut\hfill

The notation $\eta^{(j)}$ is a shorthand notation for the $j$th extension of
$\Gamma $. For the $j_1$th derivative of $u^i$ we have specifically
\begin{equation}
\eta^{i(1)}_{j_1} = D_{j_1}\eta^i - u^i_{\mbox{\sl l}}D_{j_1}\xi^{%
\mbox{\sl
l}}  \label{4.66}
\end{equation}
and for higher derivatives we can use the recursive definition
\begin{equation}
\eta^{i(k)}_{j_1j_2\ldots j_k} = D_k\eta^{i(k-1)}_{j_1j_2\ldots j_{k-1}} -
u^i_{j_1j_2\ldots j_k\mbox{\sl l}}D_{j_k}\xi^{\mbox{\sl l}}  \label{4.67}
\end{equation}
in which the terms in parentheses are not to be taken as summation indices.

\strut\hfill

The Jacobian of the transformation may be written as
\begin{eqnarray}
J\left( \bar{x},x\right) &=&\left\vert \displaystyle{\frac{\partial \bar{x}%
^{i}}{\partial x^{j}}}\right\vert  \nonumber \\
&=&\left\vert \delta _{ij}+\varepsilon D_{j}\xi ^{i}+O\left( \varepsilon
^{2}\right) \right\vert  \nonumber \\
&=&1+\varepsilon D_{j}\xi ^{j}+O\left( \varepsilon ^{2}\right) .
\label{4.68}
\end{eqnarray}%
We now can write (\ref{4.65}) as
\begin{equation}
\bar{A}=\int_{\Omega }\left\{ L+\varepsilon \left[ \varepsilon LD_{j}\xi
^{i}+\xi ^{j}\displaystyle{\frac{\partial L}{\partial x_{j}}}+\eta
_{j_{1}j_{2}\ldots j_{k}}^{i(k)}\displaystyle{\frac{\partial L}{\partial
u_{j_{1}j_{2}\ldots j_{k}}^{i}}}\right] \right\} \mbox{\rm d}x+\int_{\delta
\Omega }L\mbox{\rm d}x+O\left( \varepsilon ^{2}\right) .  \label{4.69}
\end{equation}%
Because the transformation is infinitesimal, to the first order in the
infinitesimal parameter, $\varepsilon $, the integral over $\delta \Omega $
can be written as
\begin{eqnarray}
\int_{\delta \Omega }L\mbox{\rm d}x &=&\varepsilon \int_{\partial \Omega }L%
\mbox{\rm d}\sigma  \nonumber \\
&=&-\int_{\Omega }D_{j}F^{j}\mbox{\rm d}x,  \label{4.70}
\end{eqnarray}%
where $\mathbf{F}$ is an as yet arbitrary function. The requirement that the
Action Integral be invariant under the infinitesimal transformation now
gives
\begin{equation}
D_{j}F_{j}=LD_{j}\xi ^{j}+\xi ^{j}\displaystyle{\frac{\partial L}{\partial
x_{j}}}+\eta _{j_{1}j_{2}\ldots j_{k}}^{i(k)}\displaystyle{\frac{\partial L}{%
\partial u_{j_{1}j_{2}\ldots j_{k}}^{i}}}.  \label{4.71}
\end{equation}%
This is the condition for the existence of a Noether symmetry for the
Lagrangian. We recall that the Variational Principle was not used in the
derivation of (\ref{4.71}) and so the Noether symmetry exists for all
possible curves in the phase space and not only the trajectory for which the
Action Integral takes a stationary value.\hfill

To obtain a conservation law corresponding to a given Noether symmetry we
manipulate (\ref{4.71}) taking cognizance of the Euler-Lagrange Equations.
As
\begin{equation}
\xi ^{i}D_{j}L=\xi ^{j}\left( \displaystyle{\frac{\partial L}{\partial x_{j}}%
}+D_{j}u_{j_{1}j_{2}\ldots j_{k}}^{i}\displaystyle{\frac{\partial L}{%
\partial u_{j_{1}j_{2}\ldots j_{k}}^{i}}}\right) ,
\end{equation}%
we may write (\ref{4.71}) as
\begin{eqnarray}
D_{j}\left[ F_{j}-L\xi ^{j}\right]  &=&\left[ \eta _{j_{1}j_{2}\ldots
j_{k}}^{i(k)}-\xi ^{j}D_{j}u_{j_{1}j_{2}\ldots j_{k}}^{i}\right] %
\displaystyle{\frac{\partial L}{\partial u_{j_{1}j_{2}\ldots j_{k}}^{i}}}
\nonumber \\
&=&\left( \eta ^{i}-\xi ^{j}D_{j}u^{i}\right) \displaystyle{\frac{\partial L%
}{\partial u^{i}}}+\sum_{k=1}^{p}\left[ \eta _{j_{1}j_{2}{\ldots }%
j_{k}}^{i(k)}-\xi ^{j}D_{j}u_{j_{1}j_{2}\ldots j_{k}}^{i}\right] %
\displaystyle{\frac{\partial L}{\partial u_{j_{1}j_{2}\ldots j_{k}}^{i}}}
\nonumber \\
&=&-\left( \eta ^{i}-\xi ^{j}D_{j}u^{i}\right)
(-1)^{k}D_{j_{1}}D_{j_{2}}\ldots D_{j_{k}}\displaystyle{\frac{\partial L}{%
\partial u_{j_{1}j_{2}\ldots j_{k}}^{i}}}  \nonumber \\
&&\quad +\sum_{k=1}^{p}\left[ \eta _{j_{1}j_{2}\ldots j_{k}}^{i(k)}-\xi
^{j}D_{j}u_{j_{1}j_{2}\ldots j_{k}}^{i}\right] \displaystyle{\frac{\partial L%
}{\partial u_{j_{1}j_{2}\ldots j_{k}}^{i}}}  \label{4.72}
\end{eqnarray}%
in the second line of which we have separated the first term from the
summation and used the Euler-Lagrange Equation. We may rewrite the first
term as
\begin{eqnarray*}
D_{j_{k}}\left[ \left( \eta ^{i}-\xi ^{j}D_{j}u^{i}\right)
(-1)^{k}D_{j_{1}}D_{j_{2}}\ldots D_{j_{k-1}}\displaystyle{\frac{\partial L}{%
\partial u_{j_{1}j_{2}\ldots j_{k}}^{i}}}\right]  && \\
\quad -\left( D_{j_{k}}\eta ^{i}-\left( D_{j_{k}}\xi ^{j}\right)
D_{j_{k}}u^{i}-\xi ^{j}D_{jj_{k}}u^{i}\right)
(-1)^{k}D_{j_{1}}D_{j_{2}}\ldots D_{j_{k-1}}\displaystyle{\frac{\partial L}{%
\partial u_{j_{1}j_{2}\ldots j_{k}}^{i}}}. &&
\end{eqnarray*}

The first term, being a divergence, can be moved to the left side (after
replacing the repeated index $j_{k}$ with $j$). We observe that the second
term may be written as (\ref{4.66})
\begin{eqnarray}
&&-\left( \eta _{j_{k}}^{i(1)}-\xi ^{j}u_{jj_{k}}^{i}\right)
(-1)^{k}D_{j_{1}}D_{j_{2}}\ldots D_{j_{k-1}}\displaystyle{\frac{\partial L}{%
\partial u_{j_{1}j_{2}\ldots j_{k}}^{i}}}  \nonumber \\
&\Leftrightarrow &-D_{j_{k-1}}\left[ \left( \eta _{j_{k}}^{i(1)}-\xi
^{j}u_{jj_{k}}^{i}\right) (-1)^{k}D_{j_{1}}D_{j_{2}}\ldots D_{j_{k-2}}%
\displaystyle{\frac{\partial L}{\partial u_{j_{1}j_{2}\ldots j_{k}}^{i}}}%
\right]   \nonumber \\
&&\quad +\left[ D_{j_{k-1}}\left( \eta _{j_{k}}^{i(1)}-\xi
^{j}u_{jj_{k}}^{i}\right) \right] (-1)^{k}D_{j_{1}}D_{j_{2}}\ldots
D_{j_{k-2}}\displaystyle{\frac{\partial L}{\partial u_{j_{1}j_{2}\ldots
j_{k}}^{i}}}  \nonumber \\
&\Leftrightarrow &-D_{j_{k-1}}\left[ \left( \eta _{j_{k}}^{i(1)}-\xi
^{j}u_{jj_{k}}^{i}\right) (-1)^{k}D_{j_{1}}D_{j_{2}}\ldots D_{j_{k-2}}%
\displaystyle{\frac{\partial L}{\partial u_{j_{1}j_{2}\ldots j_{k}}^{i}}}%
\right]   \nonumber \\
&&\quad +\left( \eta _{j_{k-1}j_{k}}^{i(2)}-\xi
^{j}u_{jj_{k-1}j_{k}}^{i}\right) (-1)^{k}D_{j_{1}}D_{j_{2}}\ldots D_{j_{k-2}}%
\displaystyle{\frac{\partial L}{\partial u_{j_{1}j_{2}\ldots j_{k}}^{i}}}
\label{4.74}
\end{eqnarray}%
in which we see the same process repeated. Eventually all terms can be
included with the divergence and we have the conservation law
\begin{equation}
D_{j}\left\{ F_{j}-L\xi ^{j}-\left( \eta _{jk{\ldots }{j_{k-{l}+1}}}^{i({l}%
)}-\xi ^{m}u_{mjk{\ldots }{j_{k-{l}+1}}}^{i}\right)
(-1)^{k}D_{j_{1}}D_{j_{2}}\ldots D_{j_{k-{l}}}\displaystyle{\frac{\partial L%
}{\partial u_{j_{1}j_{2}\ldots j_{k}}^{i}}}\right\} =0.  \label{4.75}
\end{equation}%
The relations (\ref{4.71}) and (\ref{4.75}) constitute Noether's Theorem for
Hamilton's Principle.\hfill

\section{Noether's Theorem: simpler form}

The original statement of Noether's Theorem was in terms of infinitesimal
transformations depending upon dependent and independent variables and the
derivatives of the former. Thus the theorem was stated in terms of
generalised symmetries \textit{ab initio}. The complexity of the
calculations for even a system a moderate number of variables and
derivatives of only low order in the coefficient functions is difficult to
comprehend and the thought of hand calculations depressing. We have already
mentioned that one is advised to calculate generalised Lie symmetries for
the corresponding Euler-Lagrange equation using some package and then to
check whether there exists an $\mathbf{F}$ such that (\ref{4.71}) is
satisfied for the Lie symmetries obtained. Even this can be a nontrivial
task. Fortunately there exists a theoretical simplification, presented by
Boyer in 1967 \cite{Boyer 67 a}, which reduces the amount of computation
considerably. The basic result is that under the set of generalised
symmetries
\[
\Gamma =\xi ^{i}\partial _{x^{i}}+\eta ^{i}\partial _{u^{i}},
\]%
where the $\xi ^{i}$ and $\eta ^{i}$ are functions of $u$, $x$ and the
derivatives of $u$ with respect to $x$, and
\[
\Gamma =\bar{\eta}^{i}\partial _{u^{i}},\quad \bar{\eta}^{i}=\eta
^{i}-u_{j}^{i}\xi ^{j}
\]%
one obtains the same results \cite{katzin1,katzin2}.

This enables (\ref{4.71}) and (\ref{4.75}) to be written without the
coefficient functions $\xi ^{i}$. This is a direct generalization of the
result for a first-order Lagrangian in one independent variable. One simply
must ensure that generality is not lost by allowing for a sufficient
generality in the dependence of the $\eta ^{i}$ upon the derivatives of the
dependent variables. The only caveat one should bear in mind is that the
physical or geometric interpretation of a symmetry may be impaired if the
symmetry is given in a form which is not its natural form. This does raise
the question of what is the `natural' form of a symmetry. It does not
provide the beginnings of an answer. It would appear that the natural form
is often determined by the eye of the beholder, \textit{cf} \cite{Moyo 00 a}%
.\hfill

The proof of the existence of equivalence classes of generalised
transformation depends upon the fact that two transformations can produce
the same effect upon a function.

\section{Conclusions}

In this review article we perform a detailed discussion on the formulation
of Noether's theorems and on its various generalizations. More specifically
we discuss that in the original presentation of Noether's work \cite{Noether
18 a} the dependence of the coefficient functions of the infinitesimal
transformation can be upon the derivatives of the dependent variables.
Consequently, a series of generalizations on Noether' theorem, like hidden
symmetries, generalized symmetries etc., are all included on the original
work of Noether. That specific point and that the boundary function on the
Action Integral can include higher-order derivatives of the dependent
variables were the main subjects of discussion for this work. Our aim was to
recover to the audience that generality which has been lost after texts, for
instance Courant, Hilbert, Rund and many others, where they identify as
Noether symmetries only the point transformations. The discussion has been
performed for ordinary and partial differential equations, while the
corresponding conservation laws/flows are given in each case. \hfill

\bigskip
\textit{AP acknowledges the financial support of FONDECYT grant no. 3160121.  PGLL
Thanks the Durban University of Technology, the University of KwaZulu-Natal
and the National Research Foundation of South Africa for support. AKH  expresses grateful thanks to UGC (India), NFSC, Award No.
F1-17.1/201718/RGNF-2017-18-SC-ORI-39488 for financial support}

\end{document}